\documentclass[12pt,a4paper]{article}

\usepackage{jheppub}

\def\al{\alpha}

\def\d{\partial}
\def\l{\left(}
\def\r{\right)}

\newcommand{\be}{\begin{equation}}
\newcommand{\ee}{\end{equation}}
\newcommand{\bea}{\begin{eqnarray}}
\newcommand{\eea}{\end{eqnarray}}
\newcommand{\bg}{\begin{gather}}
\newcommand{\eg}{\end{gather}}
\newcommand{\bseq}{\begin{subequations}}
\newcommand{\eseq}{\end{subequations}}

\def\half{\frac{1}{2}}

\newcommand{\bx}{\ensuremath{\mathbf{x}}}

\title{Pulsar timing signal from ultralight scalar dark matter}

\author[a]{Andrei Khmelnitsky}

\author[b,c]{and Valery Rubakov}

\affiliation[a]{Arnold Sommerfeld Center for Theoretical Physics,\\
Ludwig Maximilians University,\\
Theresienstr. 37, 80333 Munich, Germany}

\affiliation[b]{Institute for Nuclear Research of the Russian Academy of Sciences,\\
60th October Anniversary Prospect, 7a, 117312 Moscow, Russia}

\affiliation[c]{Department of Particle Physics and Cosmology, Physics Faculty,\\
M.V. Lomonosov Moscow State University,\\
Vorobjevy Gory, 119991, Moscow, Russia}

\abstract{An ultralight free scalar field with mass around $10^{-23} - 10^{-22}$ eV is a viable dark mater candidate, which can help to resolve some of the issues of the cold dark matter on sub-galactic scales. We consider the gravitational field of the galactic halo composed out of such dark matter. The scalar field has oscillating in time pressure, which induces oscillations of gravitational potential with amplitude of the order of $10^{-15}$ and frequency in the nanohertz range. This frequency is in the range of pulsar timing array observations. We estimate the magnitude of the pulse arrival time residuals induced by the oscillating gravitational potential. We find that for a range of dark matter masses, the scalar field dark matter signal is comparable to the stochastic gravitational wave signal and can be detected by the planned SKA pulsar timing array experiment.}
\subheader{LMU-ASC 66/13\\INR-TH-2013-26}
\notoc

\begin{document}

\maketitle\newpage

\section{Introduction and summary}

The standard $\Lambda$CDM cosmological model is very successful in describing properties of our Universe on large scales. However, there is a persisting mismatch between the observed properties of the structures on sub-galactic scales and the results of the structure formation simulations in cold dark matter (see e.g.~\cite{Primack} for a review). All known problems of CDM are different faces of the fact that the structures in CDM model are overabundant on sub-galactic scales. It is believed that these issues may find their explanation in complex astrophysical phenomena, which are not yet properly taken into account in the CDM simulations. Nevertheless, they provide a motivation to study dark matter models, in which the structure formation on sub-galactic scales is suppressed in comparison with the CDM, so-called \emph{warm} dark matter models, see e.g.~\cite{WDM}.

In this paper we consider an ultralight free scalar field dark matter. As discussed, e.g., in~\cite{Hu:2000ke} (who build upon earlier works), 
a scalar field of mass $m = 10^{-23} - 10^{-22}\,\text{eV}$ can serve as warm dark matter. This dark matter candidate is discussed in a number of papers, see e.g.~\cite{SFDM} for a recent review and references. An even lighter scalar field would behave as \emph{hot} dark matter and would fail to form observed structures, unless it constitutes a subdominant fraction of all dark matter~\cite{Amendola}. A heavier scalar field is a legitimate \emph{cold} dark matter candidate.  

Ultra-light non-interacting dark matter inside galactic halos is well described by a classical scalar field that oscillates with frequency $m$. It behaves as  perfect fluid with oscillating pressure, which averages to zero on the time scales greater than the oscillation period, and is usually treated as a pressureless dust. However, it seems to be overlooked that the oscillations in pressure induce oscillations in gravitational potentials, which can, in principle, be observable.

We study the effect of such oscillations on photon arrival time and show that it can be detected using the next generation of pulsar timing array observations~\cite{SKAGW}. In particular, we find that the effect on the pulsar timing is comparable to the effect of a monochromatic gravitational wave with characteristic strain
$
h_c \approx 2\cdot10^{-15} \l\frac{10^{-23}\,\text{eV}}{m}\r^2
$,
and frequency $f = 5\cdot10^{-9}\,\text{Hz} \l\frac{m}{10^{-23}\,\text{eV}}\r$.
A signal of similar magnitude is expected from the stochastic gravitational wave background produced by massive black hole binary systems~\cite{Sesana}. 
However, unlike the latter, the signal we consider is monochromatic. 
Such a signal will be in the reach of pulsar timing measurements based on the SKA observations~\cite{SKAGW}.

\section{Time-dependent gravitational potential}

The main effect that prevents ultra-light particles to form structures on small scales is the quantum pressure caused by the Heisenberg uncertainty principle. It smoothes out all inhomogeneities in the dark matter distribution on  scales smaller than the de~Broglie wavelength $\lambda_{dB}$ of dark matter particles. Since the characteristic momentum of the dark matter particles is $k = m v$, where $v \approx 10^{-3}$ is the typical velocity in the Galaxy, we have
\be
\lambda_{dB} \equiv k^{-1} = (m v)^{-1} \approx 600\,\text{pc} \l\frac{10^{-23}\,\text{eV}}{m} \r \l \frac{10^{-3}}{v} \r \;.
\ee
Taking into account that the number density of dark matter particles in the Galaxy is given by $n = \rho_{DM}/m$, we estimate the characteristic occupation number of dark matter in the galactic halo as
\be
\frac{\Delta N}{\Delta x^3 \Delta p^3} \sim n / k^3 = \frac{\rho_{DM}}{m \,k^3} \approx 10^{96}\l\frac{\rho_{DM}}{0.3\,\text{GeV}/\text{cm}^3} \r \l\frac{10^{-23}\,\text{eV}}{m} \r \;.
\ee
Since the occupation number is huge, the dark matter in the Galaxy is in the domain of validity of the classical field theory, and can be described by a classical scalar field $\phi(\bx,t)$. Generally, such a field can be represented as a collection of plane waves of typical momentum $k$.  The frequencies are given by the corresponding energy, which in the non-relativistic limit equals $E \simeq m + m v^2/2$. Since the characteristic time scale corresponding to the second term $(\Delta \omega)^{-1} = 2/(mv^2)$ is very large (the corresponding
length scale is of order 1~Mpc), to the leading order we can set the frequency to be equal to $m$. Thus the field we are dealing with has the form
\be\label{eq:phi}
\phi (\bx,t) = A(\bx) \cos \l m\,t + \alpha (\bx) \r \;.
\ee
Energy-momentum tensor of a free scalar field is given by
\be
T_{\mu\nu} = \d_\mu\phi \,\d_\nu\phi - \half g_{\mu\nu} \l (\d\phi)^2 - m^2 \phi^2\r \;.
\ee
Plugging here the Ansatz for the field~\eqref{eq:phi}, we see that the energy density $T_{00}$ has a dominant time-independent component
\be
\rho_{DM} \equiv T_{00} = \half m^2 A^2 \;,
\ee
and an oscillating part which, however, is proportional to $(\nabla \phi )^2 \sim k^2 \phi^2$ and hence  small,
\be
\rho_{DM}^{osc} \sim \frac{k^2}{m^2} \rho_{DM} = v^2 \rho_{DM} \; .
\label{sep18-1}
\ee
 On the other hand, the dominant term of the spatial components $T_{ij}$ oscillates in time with frequency 
\[
\omega = 2 m
\]
and amplitude of the order of $\rho_{DM}$:
\be
T_{ij} = - \half m^2 A^2 \cos{(\omega\,t + 2 \alpha)}\, \delta_{ij} \equiv p(\bx,t)\,\delta_{ij} \;.
\ee
The average value of this pressure over the oscillation period is zero. Therefore, it is usually neglected in cosmological context, and a free massive scalar field is considered as pressureless dust. However, the oscillating pressure induces oscillations in the gravitational potentials, which, as we show below, can lead to observable effects. 

To find the gravitational field produced by the scalar field dark matter in the galactic halo, let us use the Newtonian gauge, in which the metric takes the following form:
\begin{equation}
ds^2 = \big(1 + 2 \Phi(\bx, t) \big) dt^2 - \big(1 - 2 \Psi(\bx, t) \big) \delta_{ij} dx^i dx^j \;.
\end{equation}
The gravitational potentials $\Psi$ and $\Phi$ can be split into  dominant time-independent parts and parts oscillating with frequencies which are multiples of $\omega$. For our purposes it is sufficient to consider only the leading oscillating contributions:
\begin{equation}\label{eq:metric}
\Psi(\bx, t) \simeq \Psi_0(\bx) + \Psi_c(\bx) \cos{(\omega t + 2 \alpha(\bx))} + \Psi_s(\bx) \sin{(\omega t + 2 \alpha(\bx))}\;,
\end{equation}
and similarly for $\Phi(\bx,t)$. The induced gravitational potentials can be found from the linearised Einstein equations. The ${00}$ component of the Einstein equations reads
\be
\Delta \Psi = 4\pi G T_{00} = 4\pi G \rho_{DM} \;,
\label{sep18-2}
\ee
and gives the Poisson equation for the time-independent part $\Psi_0(\bx)$. Since the right hand side does not contain any time dependence to the leading order, the oscillating terms $\Psi_c$ and $\Psi_s$ are suppressed with respect to $\Psi_0$. The oscillating parts can be found from the trace part of the spatial $ij$ components of the Einstein equations
\be
- 6 \ddot\Psi + 2 \Delta (\Psi - \Phi) = 8\pi G\, T_{kk} = 24\pi G\, p(\bx,t) \;.
\ee
Plugging here the decomposition~\eqref{eq:metric} for the gravitational potentials, one can separate three terms with different time dependence. The time-independent part of this equation gives the usual relation between time-independent gravitational potentials: $\Psi_0(\bx) = \Phi_0(\bx)$. The $\sin (\omega t + 2 \al)$ term gives $\Psi_s = 0$ to the leading order. By considering the $\cos (\omega t + 2 \al)$ term we find a non-trivial result for the amplitude $\Psi_c$:
\be\label{eq:psi}
\Psi_c(\bx) =  \half \pi G {A(\bx)^2} =  \pi\frac{G\rho_{DM}(\bx)}{m^2}\;.
\ee
Note that according to eq.~\eqref{sep18-2} the time-independent part is
of order $\Psi_0 \sim G\rho_{DM}/k^2$, so the oscillating part is indeed
smaller than $\Psi_0$ by a factor of $k^2/m^2 = v^2$. This is consistent with
eq.~\eqref{sep18-2} in view of the estimate \eqref{sep18-1}.

It is also possible, although less straightforward, to find the oscillating part of the potential $\Phi$. However, as we will see below, it does not contribute to the final result.

\section{Effect on the pulsar timing}

The time-dependent oscillations~\eqref{eq:metric} of the metric induce a time-dependent frequency shift and a time delay for any propagating signal. Such a delay, in principle, can be captured in the pulsar timing experiments. For this one detects the pulse arrival times for a set of nearby millisecond pulsars, for which these times in the laboratory frame can be well modelled. The main effects that contribute to the variation of signal arrival times are the motion of the laboratory frame with respect to the Solar System barycentre, the peculiar motion of the pulsar, the dispersion by the interstellar medium, and the intrinsic variation of the pulsar period (cf.~\cite{Hobbs:2009yn}). After all these effects are taken into account, the residual variation of the arrival times can be attributed to the effect of propagation in a time-dependent metric. In particular, this method is employed for gravitational wave search (see, e.g., \cite{Jenet:2006sv} and references therein).

Let us estimate the time-dependent part of the timing residuals induced by the time variation~\eqref{eq:metric} of the metric in the case when the ultralight scalar field is the major component of the galactic dark matter halo. It is useful to express the change in the arrival time of the pulse at the time $t$ as an integral of the relative change in the arrival frequency of the pulse:
\begin{equation}\label{eq:res}
\Delta t(t) = - \int_0^t \frac{\Omega(t') - \Omega_0}{\Omega_0} \,d t' \;.
\end{equation}
Here $\Omega(t)$ is the pulse arrival frequency at the detector at the moment $t$, and $\Omega_0$ is the frequency in the absence of the time variation of the metric, which coincides with the pulse emission frequency at the pulsar. The frequency shift for a signal propagating in a non-trivial metric is analogous to the Sachs--Wolfe effect for CMB photons~\cite{Sachs:1967er}. To the leading order, the relative frequency shift is proportional to the gravitational potentials $\Psi$ and $\Phi$. For a signal emitted with frequency $\Omega_0$ at the point $\mathbf{x_p}$ at the moment $t'$ and detected at the point $\mathbf{x}$ at the moment $t$ the frequency shift is given by
\begin{equation}\label{eq:SW}
\frac{\Omega(t) - \Omega_0}{\Omega_0} = \Psi(\mathbf{x},t) - \Psi(\mathbf{x_p},t') - \int_{t'}^{t} n_i \d_i\big(\Psi(\mathbf{x''},t'') + \Phi(\mathbf{x''},t'') \big) \,dt'' \;,
\end{equation}
where $n_i$ is a unit vector in the direction of the signal propagation. The integral is taken along the signal trajectory $\mathbf{x''} = \mathbf{x''}(t'')$ in the unperturbed metric. Therefore, the time of the propagation of the pulse equals $t - t' = D$, the distance to the pulsar. Most of the pulsars used in the pulsar timing measurements are located at the distances $D \gtrsim 100\,\text{pc} \gg m^{-1}$ from the Solar System~\cite{Manchester:2012za}. Hence, the integrand in~\eqref{eq:SW} is a fast oscillating function over the integration interval. Since the spatial variation of the metric occurs on scales of the order of $\lambda_{dB}$, the integral in~\eqref{eq:SW} is suppressed by a factor $k/\omega =v \sim 10^{-3}$ in comparison with the first two terms, and thus can be neglected. We see that only the potential $\Psi(\bx,t)$ contributes to the time delay, so there is no need to calculate the oscillatory part of the potential $\Phi(\bx,t)$. Since the time-independent term $\Psi_0$ in the gravitational potential induces only a time-independent frequency shift, which is not measured in the pulsar timing observations, the resulting frequency shift is given by
\begin{equation}\label{eq:domega}
\frac{\Omega(t) - \Omega_0}{\Omega_0} = \Psi_c \big( \cos{(\omega t + 2 \alpha(\mathbf{x}))} - \cos{(\omega (t - D) + 2 \alpha(\mathbf{x_p}))}  \big)\;.
\end{equation}
Plugging~\eqref{eq:domega} into \eqref{eq:res} and subtracting the average value over the period, we obtain the expression for the time-dependent part of the timing residual at the time $t$:
\begin{equation}
\Delta t(t) = \frac{2 \Psi_c}{\omega}  \sin{\l \frac{\omega D}2 + \alpha(\mathbf{x}) - \alpha(\mathbf{x_p}) \r} \, \cos{\l \omega t + \alpha(\mathbf{x}) + \alpha(\mathbf{x_p}) - \frac{\omega D}2 \r}\;.
\end{equation}
This time residual is oscillating in time with frequency $\omega$ and has the amplitude
\be\label{eq:dt}
\Delta t_{DM} = \frac{2 \Psi_c}{\omega}  \sin{\l \frac{\omega D}2 + \alpha(\mathbf{x}) - \alpha(\mathbf{x_p}) \r} \;.
\ee
It depends on the distance to the pulsar and the scalar field phase $\al$ at the locations of the pulsar and the detector.

To estimate the sensitivity of the pulsar timing array observations to the dark matter signal, let us compare~\eqref{eq:dt} with the time-dependent timing residual induced by  gravitational wave background. For a single monochromatic gravitation wave with frequency $\omega$ and characteristic strain $h_c$ the amplitude of the timing residual is given by (see e.g. \cite{Wen:2011xc}, and references therein):
\begin{equation}\label{eq:gwdt}
\Delta t_{GW} = \frac{h_c}{\omega} \sin{\l \frac{\omega D (1 - \cos{\theta})}2 \r} (1 + \cos{\theta}) \sin{(2\psi)}\;,
\end{equation}
where $\theta$ is the angle between the directions to the source of the gravitational wave and the pulsar, and $\psi$ is the polarization angle of the gravitational wave. The signals from the dark matter halo~\eqref{eq:dt} and  gravitational wave background~\eqref{eq:gwdt} can be compared by considering root mean square values of the time residuals. By averaging the square of~\eqref{eq:gwdt} over the distance to the pulsar $D$ (taking into account that $\omega D \gg1$), the direction $\theta$, and the polarization angle $\psi$ we obtain
\be
\sqrt{\langle \Delta t_{GW}^2 \rangle} = \frac1{\sqrt3} \frac{h_c}{\omega} \;.
\ee
The average of the square of~\eqref{eq:dt} over the distance to the pulsar $D$ gives
\be
\sqrt{\langle \Delta t_{DM}^2 \rangle} = \sqrt2 \frac{\Psi_c}{\omega} \;.
\ee
Therefore, the scalar field dark matter has the same effect on the pulsar timing measurements as gravitational wave background with characteristic strain
\begin{equation}\label{eq:sig}
h_c = 2\sqrt3 \,\Psi_c =  2\cdot10^{-15} \l\frac{\rho_{DM}}{0.3\,\text{GeV}/\text{cm}^3}\r \l\frac{10^{-23}\,\text{eV}}{m}\r^2\;,
\end{equation}
at frequency
\begin{equation}\label{eq:freq}
f \equiv 2\pi \omega = 5\cdot10^{-9}\,\text{Hz} \l\frac{m}{10^{-23}\,\text{eV}}\r\;.
\end{equation}

\begin{figure}
\centering
\includegraphics[width=.8\textwidth]{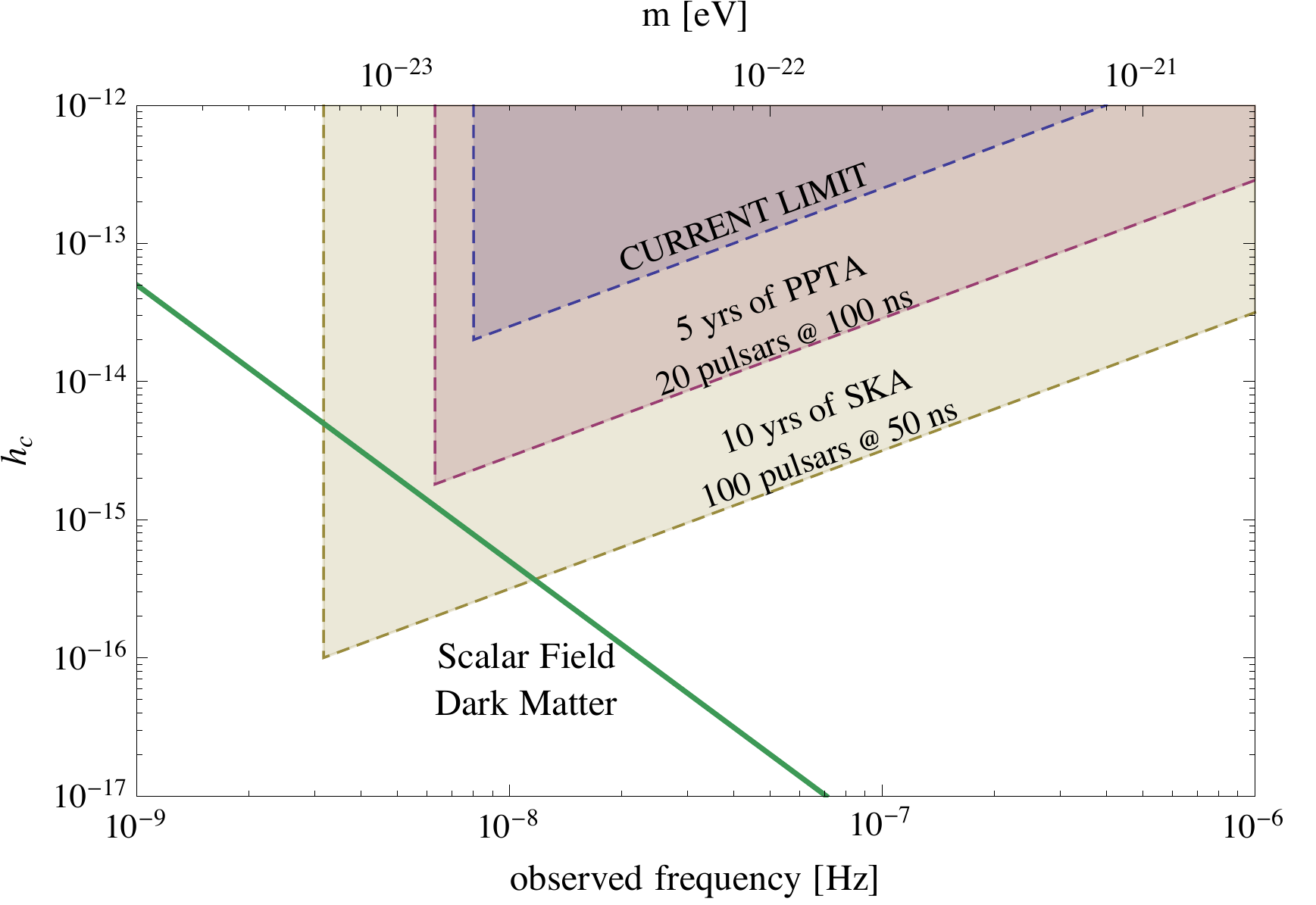}
\caption{\small  Pulsar timing signal from the scalar field dark matter~\eqref{eq:sig} for a range of scalar field masses $m$. Shaded wedges represent the estimated sensitivity of various pulsar timing array observations (adopted from~\cite{Sesana}). For masses below $10^{-23}\,\text{eV}$ the scalar field behaves like hot dark matter, and is incompatible with the observed power spectrum of density perturbations~\cite{Hu:2000ke,Marsh:2010wq}.}
\label{fig}
\end{figure}

The amplitude of the signal from the scalar field dark matter for a range of  masses $m$ is shown in Fig.~\ref{fig} together with the sensitivity curves of the pulsar timing array experiments. The sensitivities are taken from~\cite{Sesana} where three cases are considered. The current limit from the Parkes PTA~\cite{Jenet:2006sv} corresponds to $h_c \approx 2\cdot 10^{-14}$ at the frequency $f = 8\cdot 10^{-9}$ Hz. The sensitivity achievable by PPTA by monitoring 20 pulsars for 5 years with the timing precision $\delta t_{rms} = 100$ ns is estimated as $h_c \approx 2\cdot 10^{-15}$ at the frequency $f = 7\cdot 10^{-9}$ Hz. Finally, assuming that SKA will be able to monitor 100 pulsars for 10 years with the timing precision $50$ ns, the sensitivity of $h_c \approx 10^{-16}$ at the frequency $f = 3\cdot 10^{-9}$ Hz can be achieved. We see from Fig.~\ref{fig} that the scalar field dark matter signal can be observed with SKA pulsar timing array for the dark matter mass $ m \lesssim 2.3\cdot10^{-23}\,\text{eV}$.

We notice that while the signal from scalar field dark matter grows as the mass $m$ decreases, dark matter with  $m<10^{-23}\,\text{eV}$ would wash out structures on the observed scales and thus would behave as hot dark matter~(cf.~\cite{Hu:2000ke} and~\cite{Marsh:2010wq}). Hence, the interesting range of masses where the signal is detectable is fairly narrow, $10^{-23}\,\text{eV} <m \lesssim 2.3 \cdot 10^{-23}\,\text{eV}$.

To conclude, the pulsar timing signal from the scalar field dark matter is comparable to the signal expected from the stochastic gravitational wave background produced by the cosmic population of massive black hole binary systems~\cite{Sesana}. It can be detected by the pulsar timing measurements at the planned SKA telescope. The scalar field dark matter signal has two signatures that distinguish it from  stochastic gravitational wave background signal. First, the dark matter signal does not depend on the direction to the pulsar (cf.~\eqref{eq:gwdt}); second, it is monochromatic and would appear as an excess in the signal at a particular frequency related to the mass of the dark matter particle by Eq.~\eqref{eq:freq}.

\acknowledgments

The authors are indebted to Dmitry Gorbunov, Christof Wetterich, and L\={a}sma Alberte for helpful discussions. The research of AK is supported by Alexander von Humboldt Foundation. VR has been supported in part by the grants RFBR 12-02-0653 and NS-5590.2012.2.

\end{document}